\documentclass{aastex}
\usepackage{spr-astr-addons}
\usepackage{url}
\RequirePackage{color}

\begin{document}

\title{On Hoyle-Narlikar-Wheeler mechanism of vibration energy powered magneto-dipole emission of neutron stars}
\shorttitle{Vibration powered neutron star}
\shortauthors{Bastrukov, Yu, Molodtsova, Xu}

\author{Sergey Bastrukov\altaffilmark{1,2}} \and \author{Junwei Yu\altaffilmark{1}}

\and

\author{Irina Molodtsova\altaffilmark{2}} \and \author{Renxin Xu\altaffilmark{1}}

\affil{State Key Laboratory of Nuclear Physics and Technology,
School of Physics, Peking University, Beijing 100871, China}
\affil{Joint Institute for Nuclear Research, 141980 Dubna, Russia}

\altaffiltext{1}{State Key Laboratory of Nuclear Physics and Technology,
School of Physics, Peking University, Beijing 100871, China}
\altaffiltext{2}{Joint Institute for Nuclear Research, 141980 Dubna, Russia}

\begin{abstract}
We revisit the well-known Hoyle-Narlikar-Wheeler proposition that neutron star
emerging in the magnetic-flux-conserving process of core-collapse supernova can 
convert the stored energy of Alfv\'en vibrations into power of magneto-dipole radiation. We show that the necessary requirement for the energy conversion is the decay of internal magnetic field. In this case the loss of vibration energy of the star causes its vibration period, equal to period of pulsating emission, to lengthen at a rate proportional to the rate of  magnetic field decay. 
These prediction of the model 
of vibration powered neutron star are discussed in juxtaposition 
with data on pulsating emission of magnetars 
whose radiative activity is generally associated with the decay of ultra strong magnetic  field. 
\end{abstract}

\section{Introduction}
 It is generally known today that the existence of neutron stars   
 with dipole magnetic field of up to $B=10^{14}-10^{16}$ G,
 emerging in the magnetic-flux-conserving core-collapse supernova,
 has first been regarded  by Woltjer (1964). Based on this Hoyle, Narlikar and Wheeler (1964) proposed that a neutron star with such a field can generate magneto-dipole radiation by means of conversion of energy  of hydromagnetic, Alfv\'en, vibrations into power of electromagnetic emission (e.g., Pacini 2008). In recent work 
 (Bastrukov et al. 2011) such a possibility has been studied in some details 
 in the context of electromagnetic activity of magnetic white dwarfs. In this article, we present an extensive analysis of vibration-energy powered magneto-dipole radiation of a neutron star undergoing global Alfv\'en torsion seismic vibrations about axis of its dipole magnetic moment.
  In \S2, the general theory of node-free torsional Alfv\'en vibrations of perfectly conducting  degenerate stellar matter with frozen-in constant-in-time magnetic field is briefly outlined. In \S3, the theory is extended to the case of Lorentz-force-driven torsion vibrations in time-varying magnetic field. It is shown that magnetic field decay is crucial (necessary condition) to the energy conversion from Alfv\'en vibrations  to 
  magneto-dipole radiation. The predictions of the model of vibration powered  neutron star are discussed in juxtaposition with 
  available data . The results are 
  summarized in \S4 with emphasis on the relevance of the model 
  to electromagnetic activity of magnetars.

\section{Alfv\'en node-free vibrations in constant-in-time 
frozen-in magnetic field}

The starting point in the study of node-free Alfv\'en (non-compression magneto-mechanical) vibrations of a neutron star (thought of as a spherical solid mass of a perfectly conducting non-flowing continuous matter with frozen-in constant-in-time magnetic field) are equations of 
magneto-solid-mechanics  (Bastrukov et al. 2009a, 2009b, Molodtsova et al. 2010) 
  \begin{eqnarray}
  \label{e1.1}
  && \rho{\ddot {\bf u}}=\frac{1}{c}
  [\delta {\bf j}\times {\bf B}],\\
   \label{e1.1a}
  && \delta {\bf j}=\frac{c}{4\pi}[\nabla\times  
  \delta {\bf B}],
  \quad \delta {\bf B}=\nabla\times [{\bf u}
  \times {\bf B}],\\
  \label{e1.2}
  && \rho\, {\ddot {\bf u}}=\frac{1}{4\pi}
 [\nabla\times[\nabla\times [{\bf u}\times {\bf B}]]]\times {\bf B},\quad     
 \nabla\cdot {\bf u}=0.
  \end{eqnarray}
  In such vibrations the rate of differentially rotational material displacements of stellar matter is described by nodeless toroidal vector field 
   \begin{eqnarray}
   \label{e1.3}
 && {\dot {\bf u}}({\bf r},t)=
  [\mbox{\boldmath $\omega$}({\bf r},t)\times {\bf r}],\quad
  \mbox{\boldmath $\omega$}({\bf r},t)=[\nabla\chi({\bf r})]\,{\dot\alpha}(t), \\
   \label{e1.4}
  && \nabla^2\chi({\bf
  r})=0,\quad \chi({\bf r})=f_\ell(r)\,P_\ell(\cos\theta), \\
  \label{e1.4a}
 && f_\ell(r)=A_\ell\,r^{\ell}+B_\ell\,r^{-\ell-1}
 \end{eqnarray}
  which is identical to that for torsion node-free vibrations restored 
  by Hooke's elastic force (Bastrukov et al. 2007a, 2007b). In the last equation, $P_\ell(\cos\theta)$ stands 
  for Legendre polynomial of degree 
  $\ell$ specifying the overtone of toroidal $a$-mode; the amplitude $\alpha(t)$ 
  describes time evolution of vibrations (both global and locked in the crust). 
 In computing discrete spectrum of such vibrations, the magnetic field can be conveniently represented in the form: ${\bf B}({\bf r})=B\,{\bf b}({\bf r})$, where $B$ is the intensity 
  and ${\bf b}({\bf r})$ is dimensionless vector-function of magnetic field distribution 
  over the star volume. Similar representation can be used for the bulk density    
  $\rho(r)=\rho\phi(r)$, where $\rho$ is the density at the star center and  $\phi(r)$ 
  describes the radial profile of density which can be taken from computations of 
  neutron star structure relying on realistic equations of state accounting for non-uniform mass distribution in the star interior (e.g., Weber 1999).
  
  Scalar product of (\ref{e1.2}) with the following separable representation of 
  ${\bf u}({\bf r},t)={\bf a}({\bf r})\,{\alpha}(t)$ and integration over the 
  star volume leads to equation for $\alpha(t)$ having the form of 
  equation of harmonic oscillator 
  \begin{eqnarray}
  \label{e1.5}
  &&{\cal M}{\ddot \alpha}(t)+{\cal K}_m\alpha(t)=0,\, {\cal M}=\rho\,m_\ell, \, {\cal K}_m=\frac{B^2}{4\pi}\, k_\ell,\\
  \nonumber
  &&  m_\ell=\int \phi(r)\,{\bf a}({\bf r})\cdot {\bf a}({\bf r})\,d{\cal V},\\
   \nonumber
  &&  k_\ell=\int
 {\bf a}({\bf r})\cdot [{\bf b}({\bf r})\times [\nabla\times[\nabla\times [{\bf a}({\bf
 r})\times {\bf b}({\bf r})]]]]\,d{\cal V},\\
  \nonumber
 && \omega_\ell=\sqrt{\frac{{\cal K}_m}{\cal M}}=\omega_A\,\eta_\ell,\,\, \omega_A=\frac{v_A}{R}=B\sqrt{\frac{R}{3M}},\\
   \nonumber
  && \eta_\ell=\sqrt{\frac{k_\ell}{m_\ell}}\,R,\,\, v_A=\frac{B}{\sqrt{4\pi\rho}}.
 \end{eqnarray}
 The amplitude of oscillations with constant in time frequency $\omega_\ell$
  is given by  
  \begin{eqnarray}
\label{e1.8}
 \alpha(t)&=&\alpha_0\cos\,\omega t,\quad
 \alpha_0^2=\frac{2{\bar E}_A}{{\cal M}\omega^2}=
 \frac{2{\bar E}_A}{{\cal K}_m},\\
 \nonumber
 {\bar E}_{\rm A}&=&(1/2){\cal M}{\bar{\dot \alpha}^2}+
(1/2)K{\bar{\alpha^2}}\\
\label{e1.9}
 &=&(1/2){\cal M}\omega^2\alpha_0^2=(1/2)K\alpha_0^2
  \end{eqnarray}
 where over-bar stands for averaging over period of vibrations:  
 $\bar{\alpha^2}(t)=(1/2)\alpha^2_0$.
 This suggests, if all the energy $E_{\rm burst}$ of x-ray outburst goes 
  in the quake-induced vibrations and, i.e. when $E_{\rm burst}=E_A$, the amplitude $\alpha_0$    
  can be extracted from the last equation.  In the reminder of the paper we remove index $\ell$ and confine our analysis to the case of quadrupole overtone of $a$-mode, i.e., putting $\omega=\omega_{\ell=2}$. 
 \begin{figure}
\centering\
\includegraphics[width=7.0cm]{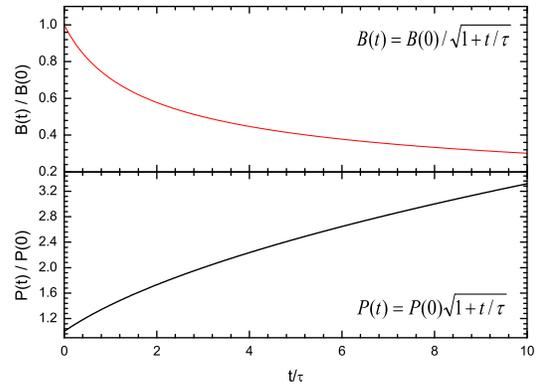}
\caption{The magnetic field decay and period elongation in a quaking 
  neutron star undergoing Lorentz-force-driven quadrupole torsion vibrations caused by conversion of vibration energy into energy of magneto-dipole emission.}
\end{figure}
  The important outcome of assumption 
  about constant in time undisturbed magnetic field is the vibration energy conservation 
  \begin{eqnarray}
\label{e1.10}
 && \frac{dE_A(t)}{dt}=0,\quad E_A(t)=\frac{ {\cal M}{\dot \alpha}^2(t)}{2}+\frac{ {\cal K}_m{\alpha}^2(t)}{2}
 \end{eqnarray}
  and also that the fundamental frequency  $\nu_A=\omega_A/2\pi$
  (where $\omega_A=v_A/R$) and the period $P_A=\nu_A^{-1}$ of global Alfv\'en oscillations
  \begin{eqnarray}
  \label{e1.11}
 \nu_A=\frac{B}{2\pi}\sqrt{\frac{R}{3M}},\quad
 P_A=\frac{2\pi}{B}\sqrt{\frac{3M}{R}}
\end{eqnarray}
remain constant in time. 
The magnitudes of basic frequency of toroidal $a$ mode in neutron stars
with magnetic fields typical to radio-pulsars $B_{12} = B/(10^{12} $ G and magnetars (soft gamma repeaters) $B_{14} = B/(10^{14}$ G are  
\begin{eqnarray}
\label{e1.12}
&&\nu_{\rm A}(B_{12}) =2.055 \cdot 10^{-3} B_{12} R^{1/2}_{6} (M/M_{\odot})^{-1/2},\\
\label{e1.13}
&&\nu_{\rm A}(B_{14})=0.2055 B_{14} R^{1/2}_{6} (M/M_{\odot})^{-1/2}
\end{eqnarray}
where $R_{6} = R/(10^{6} \; {\rm  cm}$.
It should be noted that above equations of solid-star model 
 are applicable not only to neutron stars but also for white dwarfs (Molodtsova et al. 2010; Bastrukov et al. 2010) and quark stars (Xu 2003, 2009). 
  In this work, continuing investigations reported in (Bastrukov et al. 2009a,
  2009b, 2010),  we relax the assumption about
  constant-in-time undisturbed magnetic field and examine the impact of its decay
  on quake-induced vibrations and resultant magneto-dipole radiation.

\section{Energy conversion from Alfv\'en vibrations  into 
magneto-dipole radiation}

 In the model under consideration, the decay of magnetic field (during the time of post-quake vibrational relaxation of the star) is though of as  caused by coupling of the vibrating star with outgoing material expelled by quake. With allow for dependence 
 on time of undisturbed magnetic field intensity, $B=B(t)$, the above chain of
  argument leads to equation of vibrations with the spring constant depending 
  on time: ${\cal M}{\ddot \alpha}(t)+{\cal K}_m(B(t))\alpha(t)=0$. 
 The main subject of our further analysis 
 is the impact of magnetic field decay on radiative process of neutron star undergoing Lorentz-force-driven vibrations obeying the last equation. 
It is easy to see that in this case the energy of vibrations is not conserved. The loss of vibration energy $E_A=(1/2)[{\cal M}{\dot \alpha}^2(t)+{\cal K}_m(B(t)){\alpha^2(t)}$ is determined by the rate of magnetic field decay
  \begin{eqnarray}
 \label{e2.1}
 \frac{dE_A(t)}{dt}=\frac{\alpha^2(t)}{2}\frac{d{\cal K}_m(B)}{dB}\frac{dB(t)}{dt}.
 \end{eqnarray}
 In the reminder we focus on the conversion of energy of Alfv\'en vibrations into power of its magneto-dipole radiation which is described by equation
\begin{eqnarray}
\label{e2.2}
  && \frac{dE_A(t)}{dt}=-\frac{2}{3c^3}\delta {\ddot {\bf 
  m}}^2.
  \end{eqnarray}
 Let us consider torsional magneto-mechanical 
 oscillations which are accompanied by fluctuations of total magnetic moment  $\delta {\bf m}(t)$ preserving
  its initial (in seismically quiescent state) direction ${\bf m}=m\,{\bf n}={\rm constant}$, i.e. a case when $\delta {\bf m}(t) || {\bf m}$. The total magnetic dipole moment should execute oscillations with frequency $\omega(t)$ equal to that for magneto-mechanical vibrations of  stellar matter which
are described by equation for $\alpha(t)$. This means that $\delta {\bf m}(t)$ and  $\alpha(t)$ must be subjected to equations with similar form, namely
  \begin{eqnarray}
   \label{e2.3}
  && \delta {\ddot {\bf m}}(t)+\omega^2(t)
  \delta {\bf m}(t)=0,\\
   \label{e2.4}
  && {\ddot \alpha}(t)+\omega^2(t){\alpha}(t)=0,\quad \omega^2(t)=B^2(t)
  \frac{R{\eta}^2}{3M}.
  \end{eqnarray}
  It is easy to see that equations (\ref{e2.3}) and  (\ref{e2.4}) can be reconciled if
  \begin{eqnarray}
   \label{e2.5}
  \delta {\bf m}(t)={\bf m}\,\alpha(t).
  \end{eqnarray}
 With account of all this, the equation (\ref{e2.2}) is reduced to 
 the equation of the magnetic field decay 
    \begin{eqnarray} 
    \label{e2.6}
  && \frac{dB(t)}{dt}=-\gamma\,B^3(t)\quad\to\quad B(t)=\frac{B(0)}{\sqrt{1+t/\tau}},\\
  \label{e2.7}
  && \tau=[2\gamma B^2(0)]^{-1},\quad\gamma=m^2\frac{2R\eta^2}{9M{\cal M}c^3}.
  \end{eqnarray}     
  This shows that duration of  decay, $\tau$ strongly depends upon 
  intensity of initial field $B(0)$: the larger $B(0)$, the shorter decay time $\tau$.  
   
  \subsection{Magnetic-field-decay-induced lengthening of pulse period}
  
  The most striking effect of magnetic field decay on  
  vibration-powered radiation is the lengthening of periods of pulsating magneto-
  dipole radiation, as  is demonstrated in Fig.1. The period and
   its derivative are given by
    \begin{eqnarray}
 \nonumber
 && P(t)=P(0)\,\sqrt{1+(t/\tau)},\,\,{\dot P}(t)=\frac{P(0)}{2\tau\sqrt{1+(t/\tau)}},\\
  \label{e2.8}
&& \tau=\frac{P^2(0)}{2P(t){\dot P}(t)},\quad P^2(0)=\frac{4\pi^2}{B^2(0)}\frac{3M}{R\eta^2}.
 \end{eqnarray}
 Equating two independent estimates of $\tau$, given by 
 equations (\ref{e2.7}) and   (\ref{e2.8}), one finds  that relation between 
 magnitude of total magnetic moment of the star $m$ and $P(t)$ and ${\dot P}(t)$ is 
 given by
  \begin{eqnarray}
     \label{e2.9}
  m=A \sqrt{P(t){\dot P}(t)},\quad A=\sqrt{\frac{3{\cal M}c^3}{8\pi^2}}={\rm constant}.
 \end{eqnarray} 
 For a neutron star undergoing quadrupole node-free torsion Alfv\'en vibrations 
about axis of dipole magnetic moment, the mass parameter ${\cal M}$ (represented in terms of moment of inertia $I=(2/5)MR^2$), the  magneto-mechanical stiffness ${\cal K}$, frequency $\nu$ are given by  
  \begin{eqnarray}
&& {\cal M} = \frac{9}{7}I=\frac{18}{35} MR^2,\,\, {\cal K}=\frac{6}{5}B^2\,R^3,\,\,\omega=\sqrt{\frac{\cal K}{\cal M}},\\
&& \nu(t)=\frac{\omega(t)}{2\pi}=\frac{B(t)}{2\pi}\sqrt{\frac{7R}{3M}}.
\end{eqnarray}
In these equations, the star mass $M$ and radius $R$ serve as input parameters.
For a neutron star with $M=1.4\,M_\odot$ and radius $R=10$ km, we obtain
\begin{eqnarray}
&& \nu(t)=4.6 \times 10^{-15} B\,\, (M/M_{\odot})^{-1/2}R^{1/2}_6 , \;\; {\rm Hz}\\
&& m=3.8 \times 10^{37}\,\sqrt{P(t)\,{\dot P}(t)},\,\,{\rm G}\,{\rm cm}^3.
\end{eqnarray}
These are unique to a neutron star emitting pulsed magneto-dipole radiation at the expense of energy of torsion Alfv\'en vibrations. Taking from observations $P(t)=\nu^{-1}(t)$ and ${\dot P}(t)$ one can extract intensity of magnetic field and the magnitude of total dipole magnetic moment of the star. 
Knowing $B$ and $m$, from equation {\ref{e2.7}} one can obtain the time of 
magnetic field decay $\tau$.  The practical usefulness of these estimates 
 is that they can be used as a guide in search for fingerprints of vibration 
 powered neutron stars.

     \begin{figure}
\centering
 \includegraphics[width=7.0cm]{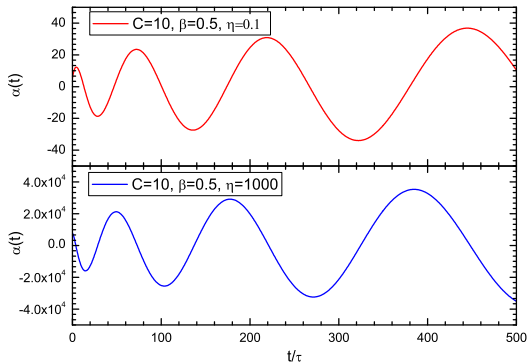}
\caption{Time evolution of the vibration amplitude, $\alpha(t)$ computed with indicated with indicated parameters $C$, $\beta$ and $\eta$.}\label{VA}
 \end{figure}

The basic conclusion of the model regarding the lengthening of vibration period 
(equal to period of electromagnetic pulse)
can be demonstrated by solution 
of equation for vibration amplitude 
\begin{eqnarray}
\label{e2.11}
 &&{\ddot \alpha}(t)+\omega^2(t)\alpha(t)=0,\quad \omega^2(t)=\frac{\omega^2(0)}{1+t/\tau}.
  \end{eqnarray}
Making us of new variable, $s= 1+t/\tau$, the above equation takes the 
form
\begin{eqnarray}
\label{e2.12}
 &&s\alpha''(s)+\beta^2\alpha(s)=0,\quad \beta^2=\omega^2(0)\tau^2
  \end{eqnarray}
 whose analytic solution reads (e.g., Polyanin and Zaitsev, 2004)
\begin{eqnarray}
\label{e2.13}
 &&\alpha(s)=s^{1/2}\{C_1\,J_1(2\beta s^{1/2})+C_2Y_1(2\beta s^{1/2})\}
  \end{eqnarray}
 where $J_1(2\beta s^{1/2})$ and $Y_1(2\beta s^{1/2})$ are Bessel functions  (Abramowitz and Stegun 1972). The arbitrary constants  $C_1$ and $C_2$ can be eliminated  from the following boundary conditions
 $\alpha(t=0)=\alpha_0,\quad \alpha(t=\tau)=0$
 where $\alpha_0$ the above defined amplitude at the initial state of vibrations (before magnetic field decay). 
 As a result, the solution of (\ref{e2.12}) can be represented in the form
 \begin{eqnarray}
\label{e2.14}
 &&\alpha(t)=C\,[1+(t/\tau)]^{1/2}\\
 \nonumber
 && \times \{J_1(2\beta\,[1+(t/\tau)]^{1/2})-\eta\,Y_1
 (2\beta\,[1+(t/\tau)]^{1/2})\},\\
 \nonumber
 && \eta=\frac{J_1(z(\tau))}{Y_1(z(\tau))},\quad C=\alpha_0[J_1(z(0))-\eta\,Y_1(z(0))]^{-1}.
  \end{eqnarray}
 The lengthening of vibration period is illustrated in Fig. 2. in which 
 the vibration amplitude is plotted as a function of $x=t/\tau$ with fixed value of $C$ and different values of parameters $\eta$ and $\beta$, respectively. This figure 
shows that $\eta$ is the parameter regulating magnitude of amplitude: the larger  $\eta$ the higher the amplitude. However, this parameter does not affect the rate of period lengthening. Both, the elongation rate of vibration
period and magnitude of vibration amplitude are highly sensitive to parameter $\beta$.

It worth emphasizing that the magnetic-field-decay-induced loss of vibration energy is substantially different from the vibration energy dissipation caused by shear viscosity of matter resulting in heating of stellar material.
The characteristic feature of this latter mechanism of vibration energy conversion into the heat (i.e., into the energy of non-coherent
electromagnetic emission responsible for the formation of photosphere of the star)  is that the frequency and period of vibrations are the same
as that in the case of viscous-free vibrations (Bastrukov et al. 2010).
However, it is no longer so in the case under consideration. It follows from
   above that the magnetic field decay resulting in the loss of total energy
   of Alfv\'en vibrations of the star causes its vibration period to lengthen at a rate
   proportional to the rate of magnetic field decay.

  \subsection{Comparison with model of rotation-energy powered neutron star}

  All the above shows that in the model of vibration-energy powered
  magneto-dipole emission under consideration, the equation
  of magnetic field evolution is obtained in similar fashion
  as equation for the angular velocity $\Omega(t)$ does in
  the standard model of rotation-energy powered magneto-dipole radiation 
  \begin{eqnarray}
\label{ei2.1}
  && \frac{dE}{dt}=-\frac{2}{3c^3}
  \delta{\ddot {\bf  m}}^2(t),\\
 \label{ei2.2} 
 &&  E=E_R(t)=\frac{1}{2}I\,{\Omega}^2(t), \quad I=\frac{2}{5}M\,R^2
  \end{eqnarray}
One of the basic postulates of the model of rotation-energy powered emission 
is that the time evolution of $\delta {\bf m}$ is governed by the equation 
\begin{eqnarray}
\label{ei2.3}
  &&\delta {\ddot {\bf m}}(t)=[\mbox{\boldmath ${\Omega}$}(t)\times [\mbox{\boldmath ${\Omega}$}(t)\times {\bf m}]],\quad {\bf m}={\rm constant},\\
\label{ei2.4}
  && \delta {\ddot m}^2=m_{\perp}^2\Omega^4,\quad m_{\perp}=m\,\sin\theta
  \end{eqnarray}  
where $\theta$ is angle of inclination of ${\bf m}$ to $\mbox{\boldmath ${\Omega}$}(t)$. The utilized in this model interrelation between 
the undisturbed total magnetic moment ${\bf m}$ and constant magnetic field ${\bf B}$ frozen in the star presumes that the neutron star matter 
is in the state of permanent magnetization ${\bf M}$ of non-ferromagnetic type: ${\bf M}=(3/8\pi)\,{\bf B}$ (see, for instance, Landau, Lifshitz, Pitaevskii 1995 [\S 76, problem 2]), namely 
 \begin{eqnarray}
  \label{ei2.5}
  m=\int M\, d {\cal V}=\frac{1}{2}B\,R^3\quad\to\quad B=\frac{2m}{R^3}.
  \end{eqnarray}
 This estimate shows that in rotation powered neutron star, the frozen-in the star  
 magnetic field operates like a passive promoter of magneto-dipole radiation, that is,
 remain constant in the process of radiation.  
 As a result, the equation of energy conversion from rotation to 
radiation is reduced to equation for rotation frequency 
\begin{eqnarray}
  \label{ei2.6}
  &&{\dot \Omega}(t)=-K\Omega^3(t),\quad K=\frac{2m_{\perp}^2}{3Ic^3},\\
\label{ei2.7}
&& \Omega(t)=\frac{\Omega(0)}{\sqrt{1+t/\tau}},\quad \tau^{-1}=2\,K\,\Omega^2(0).
  \end{eqnarray} 
  Thus, in the model of vibration-energy powered emission 
  the elongation of pulse period (and, hence, the origin of ${\dot P}$) 
  is attributed to magnetic field decay, whereas in the model of rotation-energy 
  powered emission the lengthening of pulse period is ascribed to the slow down of 
  the neutron star rotation 
  (e.g., Manchester \& Taylor 1977, Lorimer \& Kramer 2004, Bisnovatyi-Kogan  2010). 

 \begin{table}
\begin{center}
\caption{Basic frequency of Alfv\'en vibrations, $\nu_A$, and the time of magnetic field  decay, $\tau$, computed with parameters of typical pulars and magnetars.}\label{TVT}
  \begin{tabular}{cccccc}
    \hline
    & $M$($M_{\odot}$)  &$ R$(km)  &B(G)&  $\nu_{\rm A}$(Hz)  & $\tau$(yr)  \\
    \hline
  &0.8 &20 &$10^{12}$ & $3.25\,10^{-3}$ & $4.53 \, 10^{10}$\\
  &1.0 & 15 & $10^{13}$ & $2.52 \, 10^{-2}$ &$2.98\,10^{7}$  \\
  &1.1 & 13& $10^{14}$ &0.22 & $7.4 \, 10^{3}$ \\
  &1.3 & 11& $10^{15}$ &1.89  &2.38  \\
     \hline
  \end{tabular}\\
\end{center}
\end{table}

\section{Summary}
 Since celebrated hypothesis of Baade-Zwicky about birth of neutron stars in core- 
 collapse supernova and Woltier-Ginzburg proposition that they should emerge as 
 sources of super strong dipole magnetic fields,  it has been realized 
 that such a field should operate as a chief promoter of magneto-dipole 
 electromagnetic emission whose energy supply can be maintained either by   
 energy of rotation or by energy of Alfv\'en vibrations (e.g., Pacini 2008). As we 
 mentioned, the idea that neutron stars can radiate like Hertzian magnetic dipole at 
 the expense of energy of hydromagnetic vibrations has first been considered  
 by Hoyle et al. (1964), but 
 without discussion of peculiarities of vibration-energy powered emission. 
 With this in mind, one of the prime goals of this work was to explore 
 some characteristic features of such emission. What is newly disclosed here 
 is that necessary condition for the energy conversion from Alfv\'en torsion vibrations  
 into magneto-dipole radiation is the decay of internal magnetic field. 
 The basic prediction of developed theory is the lengthening of periods of pulsating 
 vibration-energy powered magneto-dipole radiation which is caused by 
 decay of internal magnetic field. Taking into account that both 
 models of rotation powered and vibration powered pulsating emission of neutron 
 stars predict lengthening of pulse period it is of interest to single out 
 the potential subclass of objects whose pulsating radiation could be 
 explained by vibrations, rather than rotation. 
  In so doing, in Table 1 we present estimates 
  of $\nu_A$ and $\tau$ as functions of increasing intensity of magnetic field 
  for neutron stars with typical masses and radii. It is seen 
  that both above characteristics are quite sensitive to $M$ and $R$ and, thus, to 
  equation of state of stellar matter. For magnetic fields of 
  typical radio pulsars, $B\sim 10^{12}$ G, the computed frequency $\nu_A$ is 
  much 
  smaller than the detected frequency of pulses. This shows 
  that considered mechanism of vibration powered pulsating 
  radiation has nothing to do with pulsating emission of radio-
  pulsars. In the meantime, for neutron stars endowed with magnetic fields 
  $B\sim10^{14}$ G our estimates of $\nu_A$  fall in the realm of 
  observed frequencies of high-energy pulsating emission of soft gamma repeaters 
  (SGRs), anomalous X-ray pulsars (AXPs) and sources exhibiting similar 
  features. According to common belief, these are magnetars - highly magnetized 
  neutron stars whose radiative activity is related with magnetic field decay (e.g., 
  Woods and Thompson 2006).  As was emphasized, the decay time of magnetic 
  field (and, hence, the duration of vibration powered pulsating radiation) strongly 
  depend on the intensity of initial magnetic field of the star: the larger 
  magnetic field the shorter its decay time. Also, the magnitude of $\tau$ indicate 
  that definite conclusion about secular lengthening of pulse period of vibration 
  powered neutron star can be made only on the basis of reliable statistics which 
  demands fairly long monitoring time, namely, several years if not decades. 
  All these features of the vibration powered radiation of a neutron star are consistent 
  with data on long-term monitoring activity of AXP, 1E 2259.1+586 (2 decades, 
  1980-2000) with using Rossi X-ray Timing Explorer (RXTE) (Gavriil et al. 2004). 
  Similar trend in decreasing frequency of long-periodic pulsating emission   
  has been revealed in RXTE observations of the source XTE J1810-197 (Ibrahim et 
  al. 2004). However, the opposite tendency (increase of pulse frequency and, 
  hence, shortening of period) has been disclosed in activity of the X-ray source 
  CXO J164710.2-455216 (Muno et al. 2006) after short time of its 
  monitoring (May and July, 2005) with using of Chandra facility. 
  The most plausible reason of this discrepancy (with both, the above observations 
  and theoretical expectations of considered here models) is too 
  short time of monitoring or, in other words, the lack of reliable 
  statistics.  With all above, we conclude that vibration powered neutron 
  stars, if exist, can be revealed among the magnetars -- AXPs/SGRs - like sources 
  whose persistent X-ray luminosity, $10^{34}<L_X<10^{36}$ ergs s$^{-1}$, 
  cannot be explained by the loss of energy of slow rotation.

\section*{Acknowledgments}
This work is supported by the National Natural Science Foundation of China
(Grant Nos. 10935001, 10973002), the National Basic Research Program of
China (Grant No. 2009CB824800), and the John Templeton Foundation.


\end{document}